\documentclass[referee]{aa}
\usepackage{graphics}
\usepackage{graphicx}

\def\aj{AJ}                   
             
\def\apj{ApJ}                 
\def\apjl{ApJ}                
\def\apjs{ApJS}               
\def\ao{Appl.~Opt.}           
             
\def\aap{A\&A}

\def\mnras{MNRAS}

\def\msol{M$_{\odot}$}
\begin{document}      
%   \thesaurus{12.03.1;12.03.3;12.03.4;12.04.2;12.12.1;11.03.1}  
% 
   \title{Asteroseismology of the planet-hosting star \\ $\mu$ Arae. \\
 II. Seismic analysis.} 
% 
%   \subtitle{} 
 
        \titlerunning{Asteroseismology of $\mu$ Arae}  
 
   \author{ 
M. ~Bazot 
\inst{1},
 S.~Vauclair 
\inst{1},    
F. Bouchy
\inst{2,3},
N.C.~Santos
\inst{4,5}
}

   \institute{Laboratoire d'astrophysique de l'observatoire Midi-Pyr\'en\'ees, CNRS, UMR 5572, UPS,  14, Av. E. Belin, 
31400 Toulouse, France \and Laboratoire d'astrophysique de Marseille, Traverse du Siphon, 13013 Marseille, France \and Observatoire de Haute-Provence, 04870 Saint Michel l'Observatoire, France \and Lisbon Observatory, Tapada da Ajuda, 1349-018 Lisboa, Portugal \and Observatoire de Gen\`eve, 51 chemin des Maillettes, 1290 Sauverny, Switzerland} 
 \offprints{M. Bazot}
\mail{bazot@ast.obs-mip.fr}
   
\date{Received \rule{2.0cm}{0.01cm} ; accepted \rule{2.0cm}{0.01cm} }

\authorrunning{ Bazot et al.}
\abstract{
As most exoplanets host stars, HD 160691 (alias $\mu$ Ara) presents a metallicity excess in its spectrum
compared to stars without detected planets. This excess may be primordial, in which case the star would be completely overmetallic, 
or it may be due to accretion in the
early phases of planetary formation, in which case it
would be overmetallic only in its outer layers. As discussed in a previous paper (Bazot and Vauclair 2004), seismology can help choosing between the two scenarios.
This star was observed during eight nights with the spectrograph HARPS at La Silla Observatory. Forty three p-modes have been identified (Bouchy et al. 2005). In the present paper, we discuss the modelisation of this star. We computed stellar models iterated to present the same observable parameters
(luminosity, effective temperature, outer chemical composition) while the internal structure was different according to the two extreme assumptions : original overmetallicity or accretion. We show that in any case the seismic constraints lead to models in complete agreement with the external parameters deduced from spectroscopy and from the Hipparcos parallax ($L$ and $T_{{\rm eff}}$). We discuss the tests which may lead to a choice between the two typical scenarios. We show that the ``small separation'' seem to give a better fit for the accretion case than for the overmetallic case, but in spite of the very good data the uncertainties are still too large to conclude. We discuss the observations which would be needed to go further and solve this question.

\keywords{stars: oscillations - stars: abundances - stars: planetary systems: formation - stars: individual: HD160691}
}

\maketitle
                                                                                                                                               
\section{Introduction} 
% Since the first discovery of a planet orbiting around Peg 51 (Mayor et al. 1995), more than 130 exoplanets have been detected \footnote{see for example http://www.obspm.fr/encycl/cat1.html}. Due to the bias of the detection techniques (radial velocity or transit methods), the observed planetary systems are different from our Solar System : up to June 2004, only Jupiter-like planets orbiting close to the central star could be observed. However three low-mass planets (of order 14 to 20 earth masses) have been recently detected (Santos et al. 2004a, McArthur et al. 2004, Butler et al. 2004). The mass of the known exoplanets range from $\sim$ 17 M$_{Jup}$ to 14 M$_{\oplus}$ for the smallest one (Santos et al. 2004a). Their orbits have radii ranging from $\sim$ 0.03 AU to $\sim$ 4 AU, with a significant fraction (around 60\%) less than 0.6 AU (Udry et al. 2003).

 Since the first discovery of a planet orbiting around Peg 51 (Mayor et al. 1995), more than 130 exoplanets have been detected \footnote{see for example http://www.obspm.fr/encycl/cat1.html}. The central stars of these planetary systems appear to be overmetallic compared to the Sun, at least in their atmospheres. Their average metallicity is $\sim$ 0.2 dex larger than solar while it is about solar for stars which have no detected planets, in the solar neighborhood (Gonzalez et al. 1998, Santos et al. 2003).
 Two scenarios have been proposed to explain these high metallicities. In the first scenario, they are the result of a high initial metal content in the proto-stellar gas : in a metal-rich environment the mechanisms leading to planet formation are more efficient, which would explain the observed bias. 
In the second scenario, the overmetallicity is due to the accretion of hydrogen-poor matter during planetary formation. This matter could come from already formed planets which migrate to wards the central star (Trilling et al. 1998) and eventually fall on it (Murray et al. 2001, Murray \& Chaboyer 2002). This process leads to a metal excess in the stellar outer layers only (as it is diluted in the surface convective zone), while the former scenario implies that planet-hosting stars are completely overmetallic, down to their centers.

Bazot \& Vauclair (2004, hereafter BV04) studied the internal differences of stellar models computed with either the overmetallic or the accretion assumption, but 
iterated so as to present exactly the same observable parameters (T$_{{\rm eff}}$, $\frac{L}{L_{\odot}}$, Z) : these models can account for the same observed star, but their interiors are different. 
Moreover, they have slightly different masses (up to 20\%) while their observed parameters are the same. BV04 studied the asteroseismic signatures of these internal differences. 
 They focused on the fact that something special occurs for stars with masses around 1.1 M$_{\odot}$, as convective cores begin to develop. This is a delicate and metallicity-dependent process (Michaud et al. 2004). Indeed, it is possible that among two models representing the same observed star, one has a convective core while the other does not. 

We decided to test these preliminary investigations with the star $\mu$ Ara (HD160691, HR6585, GJ691). It is a G3IV-V star with a visual magnitude V=5.1. We obtained 8 nights in June 2004 to observe this star with the HARPS spectrograph on the 3.6-m telescope in La Silla. We first reported on a discovery of a new low-mass planet around this star, of $\sim$14 earth masses (Santos et al. 2004b). The star itself presents very clear oscillations : the analysis of the Fourier transform of the radial velocity time series allowed us to detect 43 oscillation modes of degrees l=0 to l=3 (Bouchy et al. 2005, hereafter paper I). A summary of these observations is given below, in section 2. The model computations and comparison with the observations are presented in section 3. The results are discussed in section 4 while we conclude and propose future work in section 5.

\section{Observational constraints}
\begin{figure}
\begin{center}
\includegraphics[angle=0,totalheight=\columnwidth,width=\columnwidth]{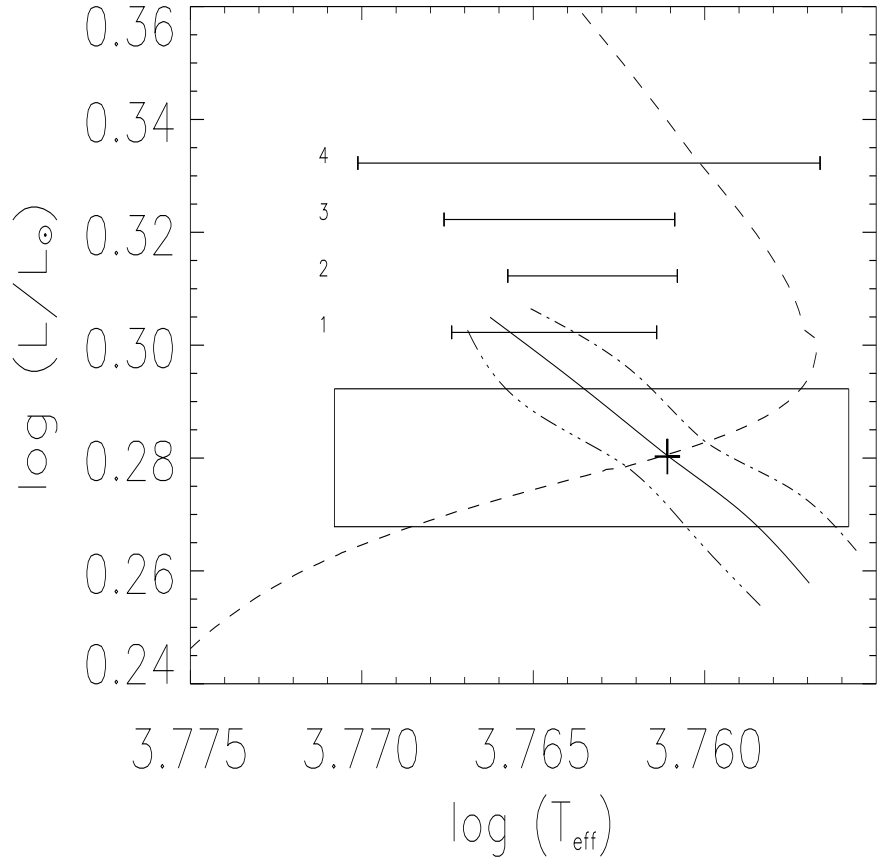}
\end{center}
\caption{This graph represents the constraints on the position of the star $\mu$ Ara in the HR diagram. Various observational constraints are shown. The error box corresponds to the luminosity derived from the Hipparcos parallax and the effective temperature measured by Bensby et al.(2003). Other temperature determinations are displayed above the box and labeled as follows : 1 - Santos et al.(2004a), 2 - Santos et al.(2004b), 3 - Laws et al.(2003) and 4 - Favata et al.(1997). An example of evolutionary track is drawn for the ``overmetallic" assumption (dashed line, see text). The asteroseismic constraints appear as three oblique lines in this HR diagram : the (solid) center line represents the models for which the large separations are exactly 90 $\mu$Hz as observed in $\mu$ Ara, while the two other lines stand for 89 $\mu$Hz (top dot-dashed line) and 91 $\mu$Hz (bottom dot-dashed line). The cross shows the place of the model that we have chosen to discuss in more details, for the comparison with the accretion case.}
\label{fig1}
\end{figure}

\begin{figure}
\begin{center}
\includegraphics[angle=0,totalheight=\columnwidth,width=\columnwidth]{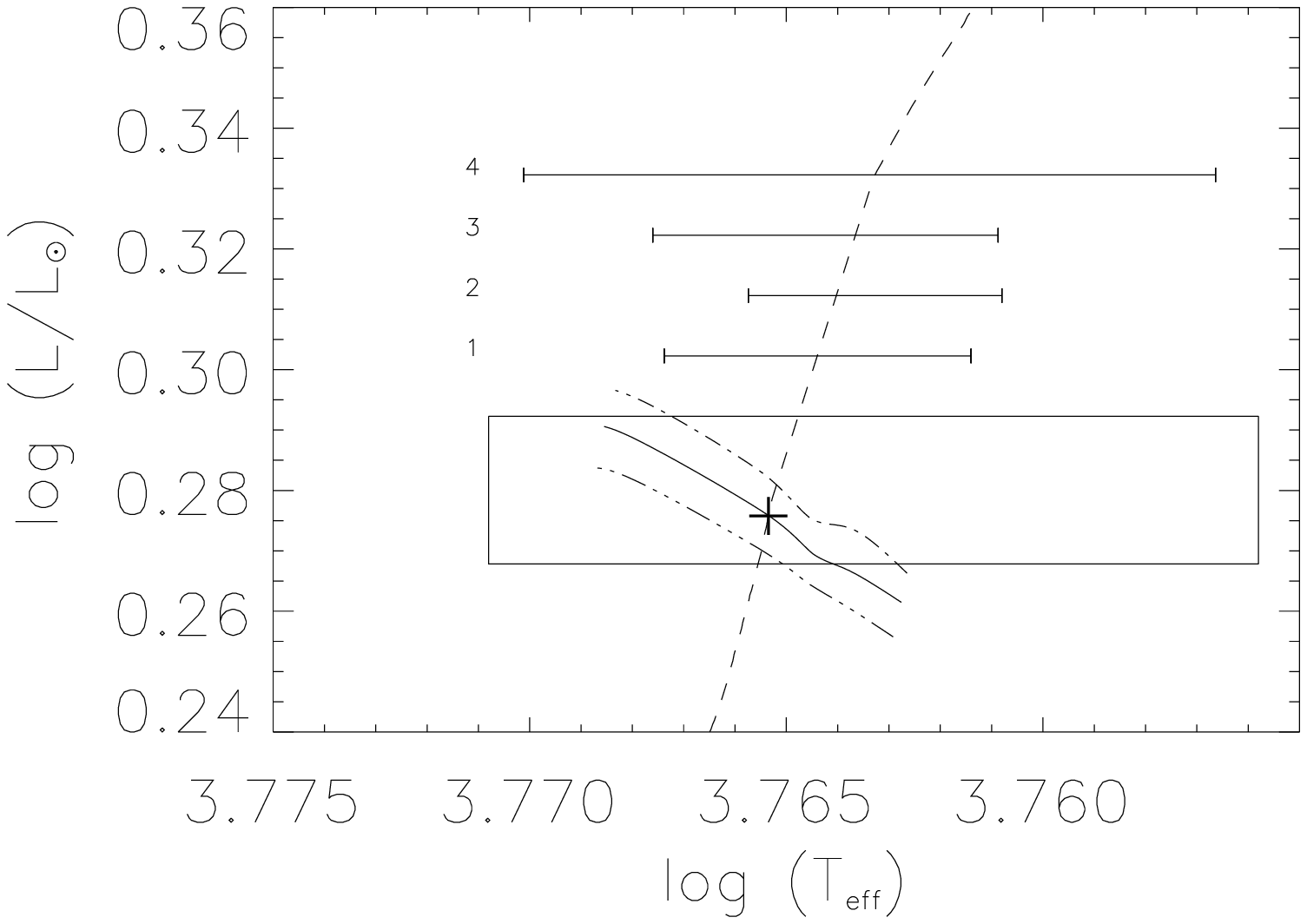}
\end{center}
\caption{Same as Fig. \ref{fig1} for the observational constraints on $\mu$ Ara. Here the evolutionary track is drawn for the ``accretion" assumption. The seismic constraints are represented for the accretion case in the same way as in Fig. \ref{fig1} for the overmetallic case.}
\label{fig2}
\end{figure}

With a visual magnitude of V=5.1 and a Hipparcos parallax of $\pi$=65.5$\pm$0.8 mas, the star $\mu$ Ara has an absolute magnitude M$_V$=4.20 and a luminosity of $\log L/L_{\odot}=$0.28$\pm$0.012.
The different values given in the literature for its effective temperature are listed in Table~\ref{tab1}. The HARPS spectrum obtained during our run (see below) was used to derive a new spectroscopic set of atmospheric parameters. From this study the values T$_{{\rm eff}}$=5813$\pm$40 K and [Fe/H]=0.32$\pm$0.05 dex were found (second line of Table \ref{tab1}). With such a metallicity, $\mu$ Ara happens to be one of the most metallic planet-hosting star (see e.g. Santos et al. 2003). As we can see in Table \ref{tab1} there is a discrepancy in effective temperature between the two oldest measurements (bottom lines) and the most recent ones which are in good agreement, around 5800 K. In Figures \ref{fig1} and \ref{fig2}, the five most recent measurements are displayed. We will see below that seismic studies of this star lead to more precise constraints on T$_{{\rm eff}}$, inside the error bars given by these spectroscopic values. We also notice that they are in good agreement with the newest effective temperatures, which are believed to be more precise than the oldest ones. The two lowest values given in Table \ref{tab1} for T$_{{\rm eff}}$ are ruled out by our results.

\begin{table}
\caption{Effective temperatures and metal abundances observed for $\mu$ Ara. [Fe/H] ratios are given in dex. The references for the given values are given in column 3, they are followed by \dag \ for spectrometric measurements and \ddag \ for photometric measurements.}
\label{tab1}
\begin{center}
\begin{tabular}{ccc} \hline
\hline
T$_{{\rm eff}} (K)$ & [Fe/H] & Reference \cr
 \hline
5798  $\pm$ 33  &  0.32 $\pm$ 0.04   & Santos et al. 2004a\dag \cr
5813  $\pm$ 40  &  0.32 $\pm$ 0.05   & Santos et al. 2004b\dag \cr 
5800  $\pm$ 100 &  0.32 $\pm$ 0.10   & Bensby et al. 2003\dag  \cr
5811  $\pm$ 45  &  0.28 $\pm$ 0.03   & Laws et al. 2003\dag    \cr
5800  $\pm$ 90  &  0.28 $\pm$ 0.04   & Favata et al. 1997\dag  \cr
5570  $\pm$ 70  &  0.39 $\pm$ 0.10   & Laird et al. 1985\ddag    \cr
5597  $\pm$ 160 &  0.41 $\pm$ 0.15   & Perrin et al. 1977\dag       \cr
 \hline
\end{tabular}
\end{center}
\end{table}

%The star $\mu$ Ara is known to harbour three exoplanets. The first discovered one (HD160691b) is a Jupiter-like planet with a mass M=1.67 M$_{Jup}$, a semi-major axis a=1.50 AU and an orbital period P=654.5 days (Butler et al. 2001, Jones et al. 2002). The second one (HD160691c) is also a long period Jupiter-like planet (M=3.1 M$_{Jup}$, a=4.17 AU, P=2986 days) whose existence has recently been confirmed (McCarthy et al. 2004). The third planet (HD160691d) was detected during our run. Its mass is similar to that of Uranus and its orbit is very close to the star (M=14 M$_{\oplus}$, a=0.09 AU, P=9.55 days) (Santos et al. 2004). 

%We used the high-precision echelle spectrograph HARPS located on the 3.6 m ESO telescope at La Silla, Chile. The observations were made in June 2004. The instrument has an average resolution of $\lambda /\Delta \lambda$=110000 and can display 71 spectral orders from 3802.5 \AA \ to 5323.1 \AA. We fixed the exposure time at 100s and collected 2105 optical spectra with an average SN ratio in the range 90-180 at 550 nm. The radial velocities were computed using a cross-correlation technique with a theoretical mask from a G2 star.

We observed $\mu$ Ara with the high-precision echelle spectrograph HARPS available on the 3.6-m ESO telescope at La Silla, Chile. As discussed in details in paper I, 43 low degree p-modes with a signal-to-noise ratio greater than 2.5 were identified. The degrees of the observed modes range from l=0 to l=3. The uncertainty on each mode frequency is 0.78 $\mu$Hz due to the duration of the time series. The eigenfrequencies for these p-modes are given in Table 2 of Paper I. The average separation between modes of consecutive order is $\Delta \nu_0$ = 90 $\mu$Hz (see paper I). It should be noted that in the further analysis we adopt a conservative point of view. Because of the different confidence levels we have on the observed frequencies (see paper I for a complete discussion), and thus on the individual large separations, we defined the region of seismic constraint as the area delimited by the iso-$\Delta \nu_0$ 89 $\mu$Hz and 91 $\mu$Hz (see Figs. \ref{fig1} and \ref{fig2})

The echelle diagram for the observed frequencies is shown in both Figures 3 and 4. Its interpretation and the comparisons with theoretical models are discussed in the following section.

\section{Models}

\subsection{Echelle diagrams}

We computed series of ``overmetallic" and ``accretion" models which could account for the observed parameters of $\mu$ Ara. We used the Toulouse-Geneva stellar evolution code, with the OPAL equation of state and opacities (Rogers \& Nayfonov 2002, Iglesias \& Rogers 1996) and the NACRE nuclear reaction rates (Angulo et al. 1999). In all our models, microscopic diffusion was included using the Paquette prescription (Paquette et al. 1986, Richard et al. 2004). The treatment of convection was done in the framework of the mixing length theory and the mixing length parameter was adjusted as in the Sun ($\alpha$ = 1.8, Richard 1998, see Richard et al. 2004).The accretion models were computed with the same assumptions as in BV04 (instantaneous fall of matter at the beginning of the main sequence and instantaneous mixing inside the convection zone). Neither extra-mixing nor overshoot were taken in account in the present computations (this will be done in future work).

Adiabatic oscillations frequencies were computed for a large number of models along the evolutionary tracks, using a recent version of the Brassard et al. (1992) code, as in BV04. In order to compare them with the modes observed in $\mu$ Ara, the frequencies were computed for angular degrees l=0 to l=3 and radial orders ranging typically from 4 to 100. The azimuthal order is always $m$=0.
For each evolutionary track, we focused on the models which fell inside the observed box in the HR diagram (cf Figures 1 and 2), namely models with luminosities and effective temperatures as given by spectroscopy. We then searched for those models, among the previous ones, which could precisely reproduce the observed echelle diagram. 

For each track, only models with computed large separations of exactly 90 $\mu$Hz could fit the observed echelle diagram. A difference of only 0.5 $\mu$Hz in the large separations would already destroy the concordance. The oblique solid lines in Figures 1 and 2 represent the location in the HR diagram of the models which correctly fit the echelle diagram and lie in the spectroscopic error box. The two other curves (dot-dashed ones) indicate models for which the large separations are respectively 89 and 91 $\mu$Hz, which do not fit the echelle diagram anymore. 

%Two examples of good fits for the echelle diagram are given in Figures 3 and 4, respectively for the overmetallic and the accretion cases. These models correspond to the crosses in Figures 1 and 2. Their characteristics are given in Table \ref{tab4}.
 We present here two models which, given our basic assumptions, fit reasonably well the observed frequencies and echelle diagram. We computed a reduced $\chi^2$ for both cases. The echelle diagrams are given in Figures 3 and 4, respectively for the overmetallic and the accretion cases. These models correspond to the crosses in Figures 1 and 2. Their characteristics are given in Table \ref{tab4}.
The accretion model is labeled AC and the overmetallic one OM. The internal distributions of elements in these models are displayed in Figure \ref{distmet} : in the AC model the overmetallicity lies only in the outer layers while it is present in the whole OM model. The initial helium mass fraction $Y_0$ is larger for the overmetallic model than for the accretion model as we assume here that the original interstellar matter was overabundant in metals and helium according to the chemical evolution of galaxies (Izotov \& Thuan 2001, see BV04 for a discussion of this assumption). From spectroscopy and photometry both stars would appear nearly identical, within the observational uncertainties, but their internal structures are quite different. The most striking fact is the presence of a convective core in the overmetallic model, with a radius of $\sim$ 0.07 R$_{\star}$ (R$_{\star}$ being the radius of the star).
\begin{figure}
\begin{center}
\includegraphics[angle=0,totalheight=\columnwidth,width=\columnwidth]{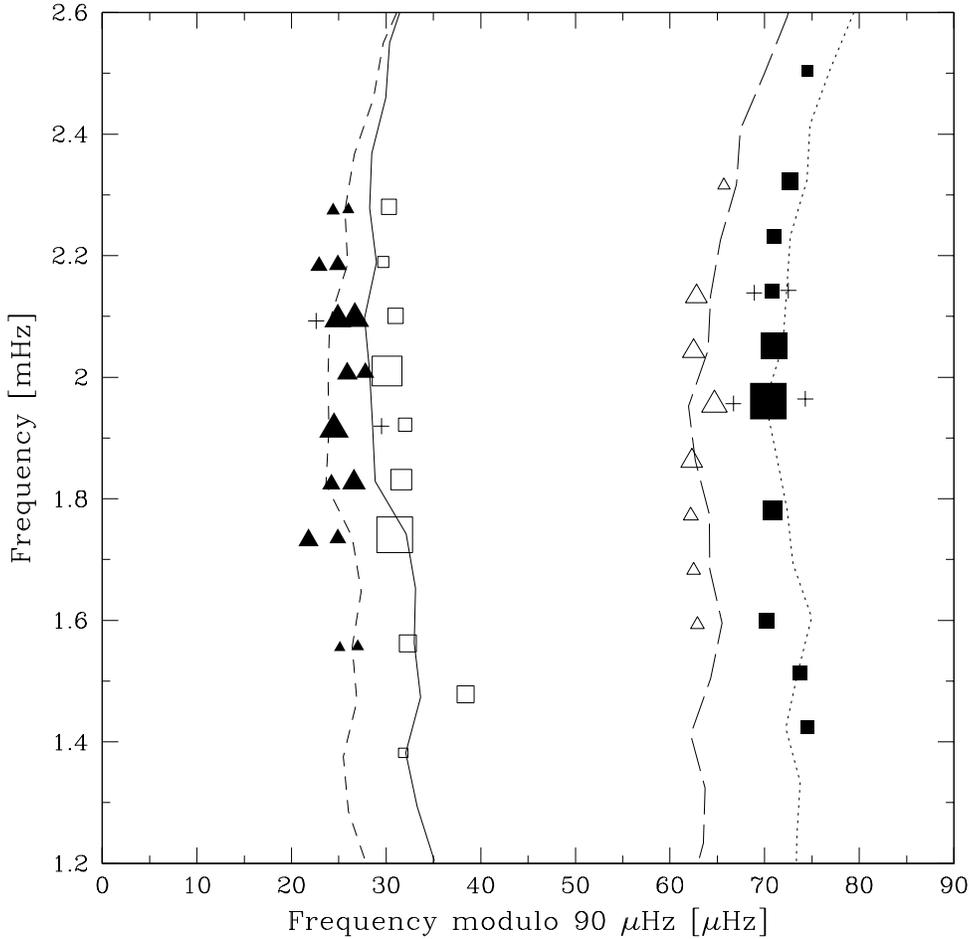}
\end{center}
\caption{ Echelle diagram for the overmetallic model. The lines represent the theoretical frequencies, the solid line is for l=0, the dotted line for l=1, the dashed line for l=2 and the long-dashed line for l=3. The symbols represent their observational counterpart. Open squares, full squares, open triangles and full triangles stand for l=0, l=1, l=2, l=3, respectively. Their size is proportionnal to the signal to noise ratio (see Paper I).}
\label{fig3}
\end{figure}

\begin{figure}
\begin{center}
\includegraphics[angle=0,totalheight=\columnwidth,width=\columnwidth]{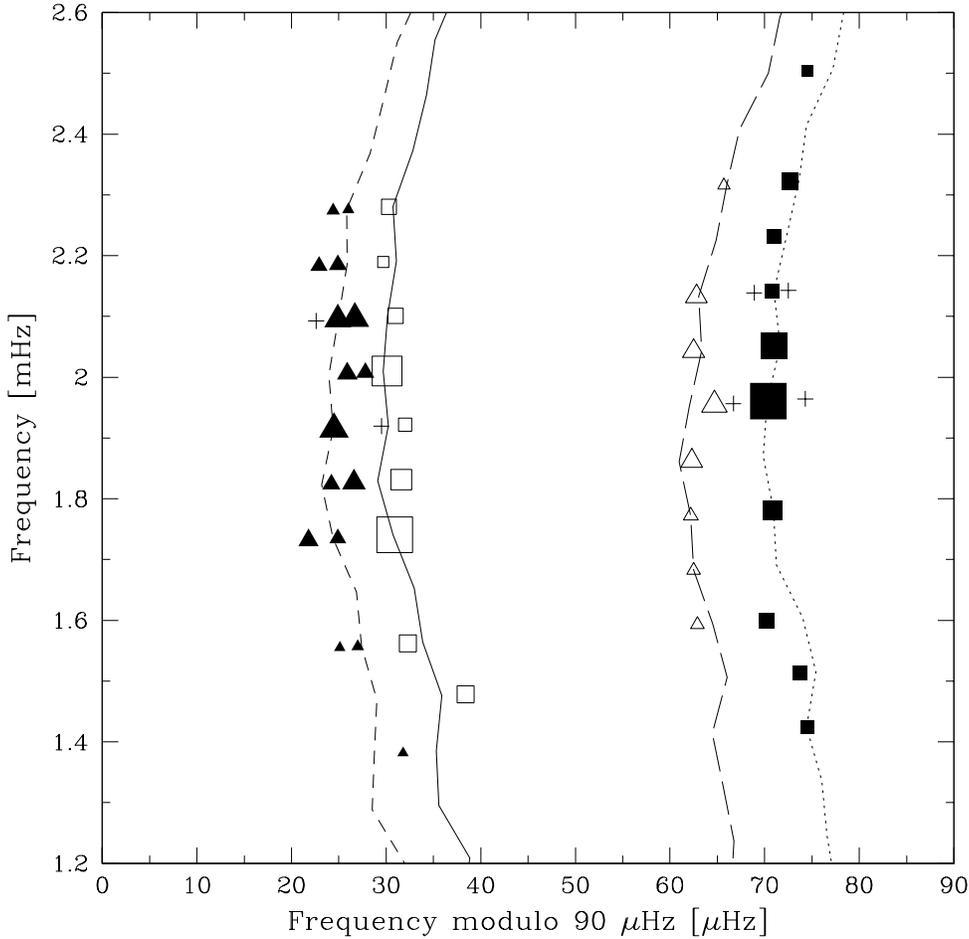}
\end{center}
\caption{Same as Fig. \ref{fig3} for the accretion model. Only the mode with the lowest frequency differs from Fig. \ref{fig3} (see text).}
\label{fig4}
\end{figure}

%{\bf The accretion scenario, with $\chi^2 = 4.3$, provides a better match for the observations than the overmetallic model, for which we obtained $\chi^2 = 7.6$. These value are still high, which questions our basic hypothesis used to fix $\alpha$ and $Y_0$. However, it is seen in the next section that a fit of the small separations gives better results.} 
%However, it can be seen from Figures 3 and 4 that the difference in the internal structures of the models cannot be detected from their echelle diagram.% Both of them fit indeed very well the observations.
 The accretion scenario, with $\chi2 = 4.3$, provides a better match for the
observations than the overmetallic model, for which we obtained $\chi2 =
7.6$. We must however keep in mind that these values are still high : it is
not possible to give a clear conclusion at the present stage.
We discuss below other possible seismic tests.

 Among these frequencies, we note that the lowest one is particularly interesting. It has been identified as a l=2 mode (see Paper I). It is in good agreement with the accretion model. However, the overmetallic model gives a slightly better fit if we consider it as being a l=0 mode. It is a good example of the uncertainty we have on the mode identification at very low signal to noise ratio.
\begin{figure}
\begin{center}
\includegraphics[angle=0,totalheight=\columnwidth,width=\columnwidth]{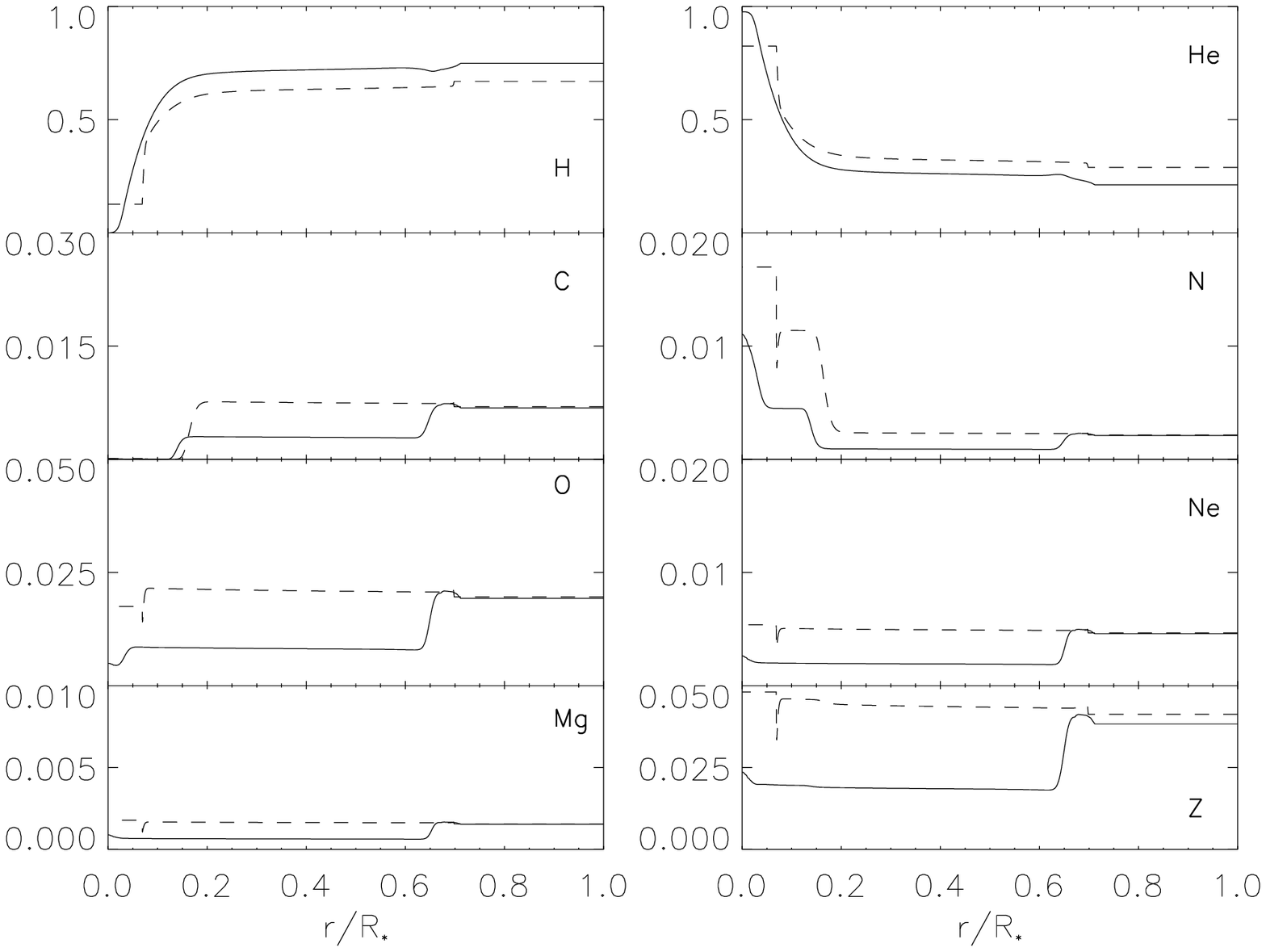}
\end{center}
\caption{Internal distribution for elements H, He, C, N, O, Ne and Mg for models OM (dashed lines) and AC (full lines). The lower right panel displays the global fraction, Z, of elements heavier than helium.}
\label{distmet}
\end{figure}

 \begin{table*}
\caption{Mass, age, initial and actual chemical compositions, radii, geometrical and acoustic, of the star (R$_{\star}$ and T$_{\star}$) and of the convective core (R$_{cc}$ and t$_{cc}$, when necessary) for stellar models which satisfy to the observational constraints for $\mu$ Ara. The model denoted AC includes accretion, the other one, denoted OM, is overmetallic. Initial mass fractions (X$_0$, Y$_0$, Z$_0$) on the main sequence are given for the core and the envelop, which has been  enhanced by pollution in the AC case. For the final model, the surface abundances are given (X, Y, Z). }
\label{tab2}
\begin{flushleft}
\begin{tabular}{ccccccccccc} \hline%\multicolumn{2}{c}{(Gyr)}\multicolumn{2}{c}{$\overline{ -0.05\ \ \ +0.05}$}
\hline
Model & M$_{\star}$ & Age  &T$_{{\rm eff}}$&L/L$_{\odot}$& \multicolumn{2}{c}{X$_0$} & \multicolumn{2}{c}{Y$_0$}  & \multicolumn{2}{c}{Z$_0$}\cr
 & (\msol) &(Gyr)  &(K)&&\multicolumn{2}{c}{$\overline{envel.\ \ \ core}$}  &\multicolumn{2}{c}{$\overline{envel.\ \ \ core}$} & \multicolumn{2}{c}{$\overline{envel.\ \ \ core}$} \cr
 \hline
AC    & 1.085  & 5.803 &5825.69& 1.887& 0.6913&0.7097  &0.2650    &0.2714  &0.0437  &0.0189    \cr 
%SM & 1.180  & 4.236 &&& 0.6260 &0.6260 &0.3297 &0.3297 &0.0433 &0.0433    \cr
OM & 1.180  & 4.251 &5769.28&1.908& 0.6696 &0.6696 &0.2891 &0.2891 &0.0413 &0.0413    \cr
 \hline\cr
\end{tabular}
\end{flushleft}
\begin{flushleft}
\begin{tabular}{ccccccc} 
\hline
\hline
X &Y &Z & R$_{\star}$& R$_{cc}$&T$_{\star}$&t$_{cc}$  \cr
 & & & (R$_{\odot}$)& (R$_{\star}$)&(s)& (s) \cr
 \hline
0.7492   &0.2124  &0.0384  &1.35 & - &5258.54& -     \cr 
0.6694 &0.2892 &0.0414&1.38 &0.07 &5240.80& 137.19   \cr
 \hline
\end{tabular}

\end{flushleft}
\end{table*}

\subsection{Small separations }

We compared the measured small separations, $\delta \nu = \delta \nu_{n,l} - \nu_{n-1,l+2}$, with the computed ones. As this quantity is very sensitive to the deep stellar interior (Tassoul 1980, Roxburgh \& Vorontsov 1994), it was proposed in BV04 as a good indicator of the presence of a convective core in stars around 1.1 M$_{\odot}$.

According to Roxburgh and Vorontsov (1994), the presence of a convective core leads to an oscillatory signal in the small separations with a period scaling approximately as $\sim 1/2t_{{\rm cc}}$, where $t_{{\rm cc}}$ is the acoustic radius of the convective core (see Table \ref{tab2}). 

%Figures \ref{fig5} and \ref{fig6} display the observed and computed small separations for the $l=0,2$ and $l=1,3$ cases. We can see an important difference between the overmetallic and the accretion cases for the $l=0,2$ small separations : in the overmetallic case they rapidly decrease at large frequencies and even become negative at some point while in the accretion case the curve is much flatter. This behavior has to be related to the echelle diagrams : for the overmetallic case (Figure 3), the theoretical lines for $l=0$ and $l=2$ cross around $\nu = 2.7 mHz$, which never happens for the accretion case. Such a crossing is in contradiction with the asymptotic theory (JCD) which may be a good representation of seismology in a radiative interior but does not hold in the presence of a convective core. It assumes in particular that the sound velocities at the turning points of the waves are identical to the central sound velocity, which is not true in this case (Figure \ref{fig7}). A complete discussion of the signatures of convective cores will be given in a forthcoming paper. Here we only compare the observed and predicted small separations in the star $\mu$ Ara. From Figures 5 and 6 we can see that the observed points seem in better fit with the accretion case than the overmetallic case. However more observations would be needed, especially for slightly larger frequencies, to be able to conclude.

\begin{figure}
\begin{center}
\includegraphics[angle=0,totalheight=\columnwidth,width=\columnwidth]{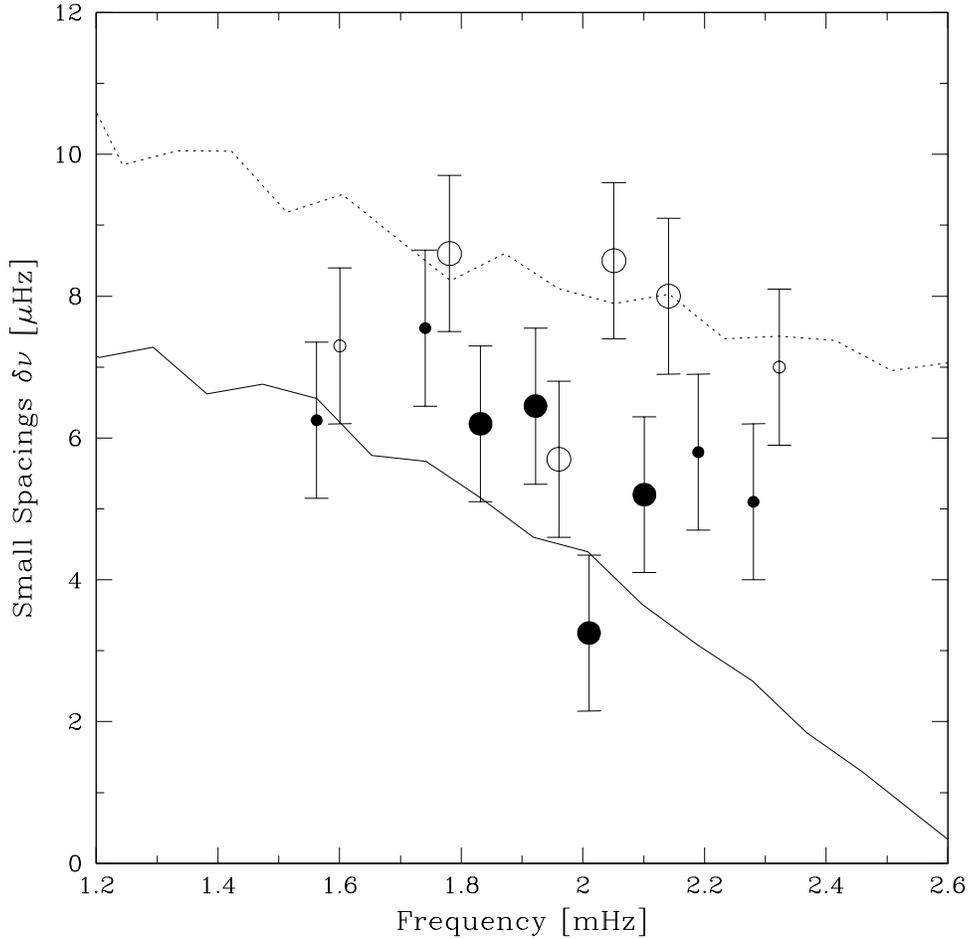}
\end{center}
\caption{ Small separations for model OM. The full and open circles are the observed values obtained for l=0-2 and l=1-3. The small dots represent values involving modes detected at a low signal to noise ratio (see Paper I). The theoretical computations are represented as a solid curve for l=0-2 and a dotted curve for l=1-3.}
\label{fig5}
\end{figure}

\begin{figure}
\begin{center}
\includegraphics[angle=0,totalheight=\columnwidth,width=\columnwidth]{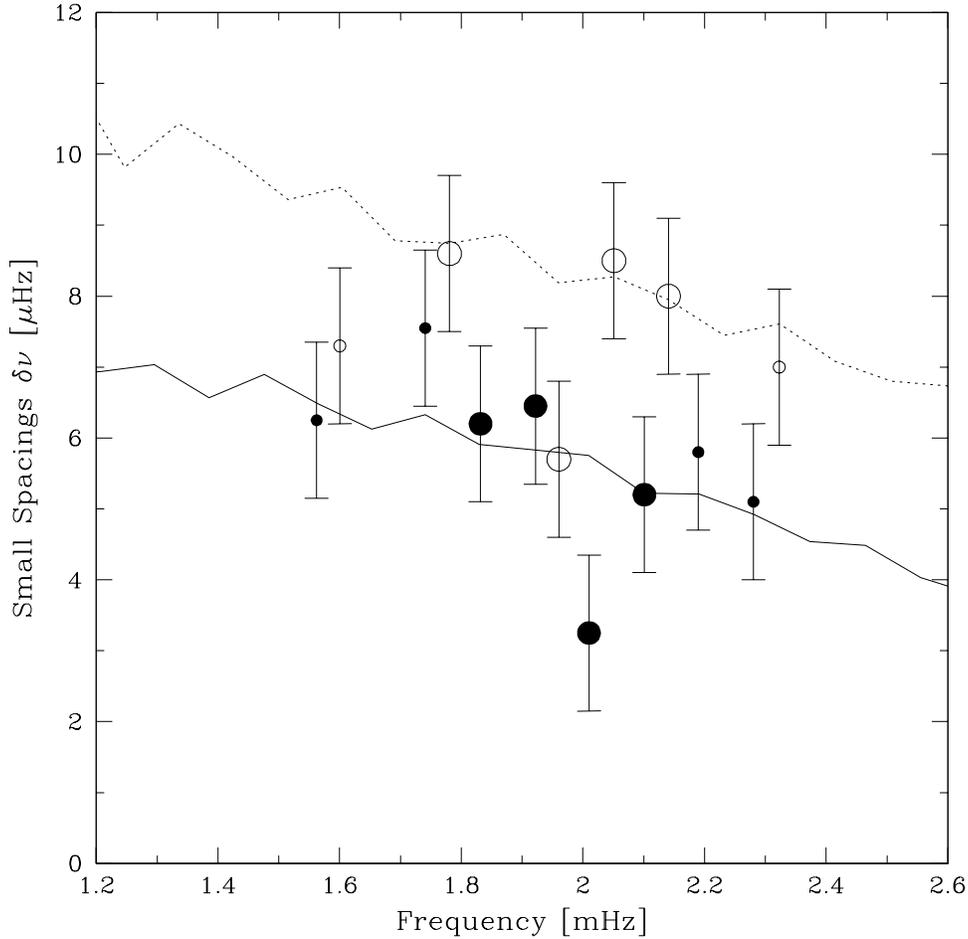}
\end{center}
\caption{Same as Fig. \ref{fig5} for model AC.}
\label{fig6}
\end{figure}

\begin{figure}
\begin{center}
\includegraphics[angle=0,totalheight=\columnwidth,width=\columnwidth]{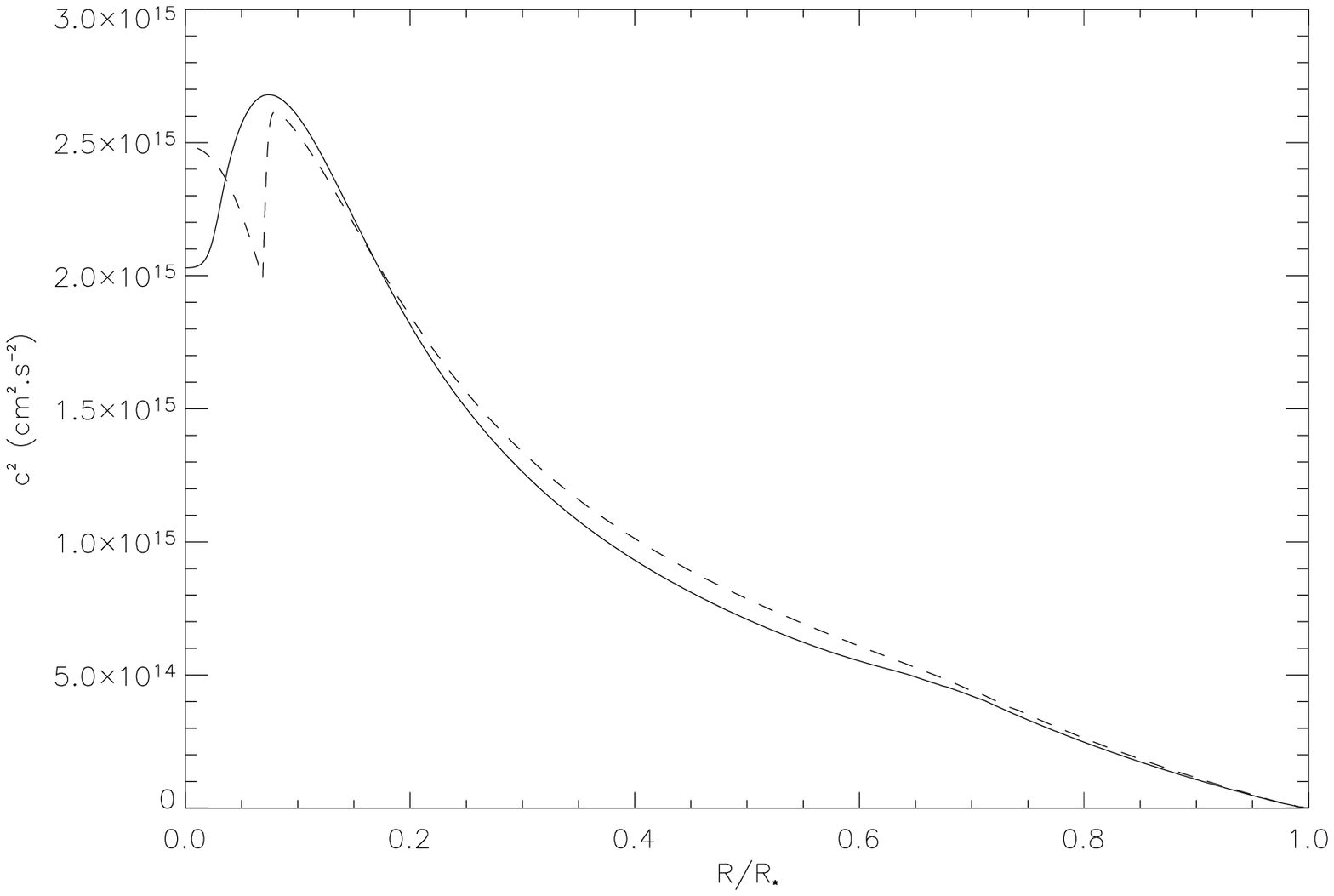}
\end{center}
\caption{Sound speed profiles inside AC (solid line) and OM (dashed line) models. The feature in the central region for the OM case compared to the AC case is due to the convective core. The sound velocity drops dramatically at its boundary and goes up again inside. The frequencies of the modes $l=0$ which progress down to the stellar center are affected by this behavior while those for which the turning point lie outside this region are not. This explains the special behavior of the corresponding lines in the echelle diagram also seen in the small separations. 
}
\label{fig7}
\end{figure}

Figures \ref{fig5} and \ref{fig6} display the observed and computed small separations for the $l=0,2$ and $l=1,3$ cases. We can see an important difference between the overmetallic and the accretion cases for the $l=0,2$ small separations. In the overmetallic case they rapidly decrease at large frequencies and even become negative at some point, while in the accretion case the curve is much flatter. This behavior has to be related to the echelle diagrams : for the overmetallic case (Figure 3), the theoretical lines for $l=0$ and $l=2$ cross around $\nu$ = 2.7 mHz, which does not happen for the accretion case. Such a crossing is in contradiction with the asymptotic theory which may be a good representation of seismology in a radiative interior but does not hold in the presence of a convective core. It assumes in particular that the sound velocities at the turning points of the waves are identical to the central sound velocity, which is not true in this case (Figure \ref{fig7}). A complete discussion of the signatures of convective cores will be given in a forthcoming paper. Here we only compare the observed and predicted small separations in the star $\mu$ Ara.

 We also computed a reduced $\chi2$ for the small separations. We obtained
$\chi2 = 1.2$ for model AC and $\chi2 = 2.2$ for model OM. These results
are better than those resulting from the fits of the frequencies (echelle
diagrams). This is partly due to the fact that the uncertainties on small
separations are $\sqrt{2}$ times larger than for individual frequencies. But
there is also a physical reason for this better fit : the small separations
are believed to be independent of the surface effects which affect the
individual frequencies (see e.g. Gabriel 1989). These effects,  which are
difficult to estimate, may deteriorate the quality of the fit of individual
frequencies while they have no influence on small separations.

The fact that model AC better matches the observations than model OM has to
be considered cautiously. We can see in Fig. \ref{fig5} that the decrease of
the theoretical small separations due to the convective core, in model OM,
becomes more important at high frequencies : this effect, which is crucial
for the choice between models, corresponds to the beginning of the
oscillations that we have discussed above. However, the three values that we
obtain for the highest observed frequencies involve modes with a signal to
noise ratio between 2.5 and 3 (small dots in Figs \ref{fig5} and \ref{fig6}). We cannot exclude
misidentifications of these modes. In such a case the corresponding small
separations would be different and the agreement between the observations
and model OM could become comparable to the one with model AC.

More observations are needed to reach a clear conclusion. It would be
particularly important in this framework to detect modes with large
frequencies, even larger than in the present work, and with the highest
possible signal-to-noise ratio, to be able to distinguish clearly between
the two scenarios.

\section{Uncertainties}
 
 Stellar models are characterized by their mass (M), an initial helium mass fraction (Y$_0$) and a mixing-length parameter ($\alpha$), and an age.

  Comparing Figs. \ref{fig1} and \ref{fig2} it is clear that the region of seismic constraint depends of the physics used in the model. The lines of constant average large separation are given for a set ($\alpha$, Y$_0$). Here $\alpha$ is adjusted on the solar models (Richard et al. 2004) and Y$_0$ is derived from the cosmic Y-Z relation (Izotov \& Thuan 2001) Changes in $\alpha$ or Y$_0$ would also lead to modifications of the location of the iso-$\Delta \nu_0$ lines in the HR diagram. To estimate this effect, we have computed models with the same input parameters as AC or OM except that we imposed a small mixing-length or initial helium mass fraction variation (only one at the time). We then selected, for given physics (i.e. accretion or overmetallicity {\it and} a given set of parameters ($\alpha$, Y$_0$)), the age for which the average large separation is 90 $\mu$Hz.

Our results are given in Table \ref{tab3}. The given variations in T$_{{\rm eff}}$ and L/L$_{\odot}$ are the differences between the values of the recomputed models and those of AC or OM. They are representative of the accuracy we can obtain with asteroseismology on $\alpha$ and Y$_0$. The changes we imposed on the mixing-length parameter and on the initial helium mass fraction correspond respectively to 3\% and 4\% of the initials values used in AC and OM. In fact, variations in Y$_0$ with such an amplitude are ruled out because the resulting position of the star in the HR diagram would not account for the spectroscopic measurements anymore. On the contrary changing $\alpha$ of 3\% has a very small influence on the star luminosity and does not affect the effective temperature in such a important way as a variation in Y$_0$ does.\\

 The problem is different for the precision obtained on the mass and the age of the star. In this case, a variation in these parameters will have no influence on the location of the iso-$\Delta \nu_0$ lines (since neither $\alpha$ nor Y$_0$ vary). For given physics, the knowledge of the effective temperature, the luminosity and the average large separation of the star helps to constrain values for its mass and its age. To estimate the precision obtained on both of these parameters, we inversely computed models including the same physics, but in which we changed either the mass or the age. Table \ref{fig4} shows the effect of a mass variation of $\pm$ 0.01 \msol \ and an age variation of $\pm$ 100 Myr (only one parameter varying at once) on T$_{{\rm eff}}$, L/L$_{\odot}$ and $\Delta \nu_0$. 

 It appears that in all cases it does not lead to major variations, neither in effective temperature nor in luminosity. In contrast, variations in $\Delta \nu_0$ are much more significant and do not agree with the seismical constraints in case of a mass change of the order of 1\%. It thus gives an upper limit of 0.01 \msol \ for the precision obtained with asteroseismology on the mass of the star. 

In case of an age shift of $\pm$ 100 Myr, the resulting models have average large separations corresponding to the limit values we used. Consequently, we expect, for given physics, an accuracy of the order of 100 Myr on the age of the star.
%% From this results, we tried to sort out a first evaluation of the precision we obtain on stellar parameters. Stellar models are caracterized by the initial helium fraction, Y$_0$, mixing-lenght-parameter, $\alpha$, age, t, and mass, M. The regions of seismic constraint provided in Fig. \ref{fig1} and \ref{fig2} correspond to a given set ($\alpha$,Y$_0$). Changing one of these parameters will change the location of the iso-$\Delta \nu_0$ lines. Taking our example models as starting points, we imposed an initial variation of the helium fraction or of the mixing-length. We then localized the position of the new model with $\Delta \nu_0$=90 $\mu$Hz in the HR diagram. Our results are displayed in Table 3. The amplitude of the variation imposed on the parameter is acceptable if the resulting variations on the effective temperature and luminosity are still in agreement with the spectroscopic constraints.
% 
%% The problem is slightly different when trying to characterize the effects of a variation of the mass or the age of the model. This time, we can still refer to the already found seismic constraints, since the couple ($\alpha$,Y$_0$) is left unchanged. We thus made the mass or the age vary (see Table 4). In this case, the amplitude of the variation imposed on the parameter is acceptable if the resulting variations on the effective temperature, luminosity {\it and} average large separation are still in agreement with the spectroscopic {\it and} the seismic constraints.
\begin{center}
\begin{table}
\begin{center}
\caption{Difference of effective temperature and luminosity between sample models, AC or OM, and new models with $\Delta \nu_0$=90 $\mu$Hz but with a variation of either the initial helium fraction or the mixing-length parameter. The effective temperatures are given in K.}
\label{tab3}
\begin{tabular}{cccccc}\hline
\hline
Model&&\multicolumn{2}{c}{$\Delta \alpha$} & \multicolumn{2}{c}{$\Delta$Y$_0$} \cr
&&\multicolumn{2}{c}{$\overline{ -0.05\ \ \ +0.05}$}  & \multicolumn{2}{c}{$\overline{-0.01\ \ \ +0.01}$}\cr
\hline
AC&$\Delta {\rm T_{eff}}$& -5.5   & +26.2   & -53.1   &  +69.8      \cr
  &$\Delta ({\rm L/L_{\odot}})$& -0.027   & +0.029   & -0.079   & +0.142       \cr
\hline
OM&$\Delta {\rm T_{eff}}$& -7.3   & +9.7   & -57.4   & +66.3 \cr
  &$\Delta ({\rm L/L}_{\odot})$& -0.009   & +0.016   & -0.073   & +0.084 \cr
\hline
\end{tabular}
\end{center}
\end{table}
\end{center}

\begin{center}
\begin{table}
\begin{center}
\caption{Variation of the effective temperature, luminosity and average large separation for models AC and OM when subjects to a small variation of mass or age. The mass are given in M$_{\odot}$, the age in Myr the effective temperatures in K and the average large separation in $\mu$Hz.}
\label{tab4}
\begin{tabular}{cccccc}\hline
\hline
Model&&\multicolumn{2}{c}{$\Delta$ M} & \multicolumn{2}{c}{$\Delta$t} \cr
&&\multicolumn{2}{c}{$\overline{ -0.01\ \ \ +0.01}$}  & \multicolumn{2}{c}{$\overline{-100\ \ \ +100}$}\cr
\hline
  &$\Delta {\rm T_{eff}}$&+3.3    &-1.5    & +1.2   &  +1.8      \cr
AC&$\Delta ({\rm L/L_{\odot}})$& +0.003   &  -0.003  & +0.007   &  -0.006      \cr
  &$\Delta {\rm (\Delta \nu_0)}$& -3.0   & +3.0   & +1.0   &   -1.0     \cr
\hline
  &$\Delta {\rm T_{eff}}$&+8.0    &-3.1    & +11.1   &   -11.1     \cr
OM&$\Delta ({\rm L/L}_{\odot})$& -0.017   & +0.017   & -0.002   & +0.002       \cr
  &$\Delta {\rm (\Delta \nu_0)}$& +3.0   & -2.0   & +1.0   & -1.0       \cr
\hline
\end{tabular}
\end{center}
\end{table}
\end{center}

\section{Discussion and conclusions.}

 For the first time stellar oscillations have been detected and an asteroseismic analysis has been performed for a planet-hosting star. With a magnitude V=5.1, $\mu$ Ara also represents the faintest solar-like star with detected p-modes. The observing run was a real success, as 43 modes could be identified, leading to a clear echelle diagram with large separations of 90 $\mu$Hz. Note that we also discovered during this run the smallest exoplanet ever observed, with a mass of 14 earth masses.

The seismic analysis of this star leads to precise constraints on its position in the HR diagram, inside the observational box obtained from spectroscopy and the Hipparcos parallax. 

The ultimate aim of this run was to try to determine whether the star is completely overmetallic or only in its outer layers. In the first case it would suggest that the star and its planets formed out of an interstellar nebula which was already metal rich, while in the second case it would mean that accretion at the beginning of planetary formation was much more active than generally thought. 

We found an interesting test to decide between these two fundamental scenarios : the overmetallic models which can account for the observable parameters of the star have a convective core while the accretion models do not. We chose two of these models to go further in the analysis, by comparing the large and the small separations. We found that the presence of the convective core in the overmetallic model presented a signature in its deviations to the asymptotic theory : the lines $l=0$ and $l=2$ cross in the echelle diagram, which does not happen in the accretion model. 

Unfortunately this signature clearly appears around frequencies of 2.5 mHz which correspond to the highest frequency we were able to detect. At first sight, the graphs which display the observed and computed small separations (Figures \ref{fig5} and \ref{fig6}) seem in favor of the accretion scenario, but we definitely need more observations to go further. It would be necessary to increase the signal to noise ratio for frequencies around 2.5 to 2.7 mHz, which lie in the crucial region where the small separations for the overmetallic and the accretion models clearly diverge. 

% This study can also help us to give a constraint on the inclination of the orbital plane of the $\mu$ Ara system. In paper I a value of R$\sin i$ = 1.04 $\pm$ R$_{\odot}$ was derived for the radius of $\mu$ Ara. For model AC and OM it leads us respectively to inclinations of {\it i} = 50.4 and {\it i} = 48.9. Thus, considering the mass of the star in both cases we can obtain estimates of the real mass of the planet HD160691d which could be around 18 M$_{\oplus}$ if the assumption of accretion is verified or 20 M$_{\oplus}$ in the framework of original overmetallicity scenario.  

%(((((((((( {\bf Discussion sur rayon de l'etoile, inclinaison et masse de la planete})))))))))))))

%(((((((((({\bf Francois : }finish with discussion about future observations : HARPS, COROT (the first planet hosting target: give its name, parameters and so on) , dome C)))))))))))))))

In spite of a large amount of observational constraints 
on the bright star $\mu$ Ara, we were not able to find a unique solution for the mass of the star and propose two possible masses 
which differ by more than 10\%. 
Further additional Doppler observations, in mono or multi-sites, 
will not easily resolve the question.  In the 
frequency range we are trying to reach, p-mode amplitudes are lower than 10 cm.s$^{-1}$ and we 
can suspect that, due to their short life time, it will be very 
difficult to extract them from granulation noise. 
Future interferometric determinations of the radius of $\mu$ Ara 
could help choosing one of the two scenarios, in case it can reach 
an accuracy better than 1\%. 
Our analysis clearly illustrates the difficulties of characterizing 
the internal structure of stars even with a large number of 
identified p-modes. 

\begin{acknowledgements}
                                                                                
The authors thank S. Charpinet for his very constructive remarks. We are grateful to ESO staff support at the 3.6-m telescope. S.V. acknowledges a grant from Institut Universitaire de France. N.C.S. would like to thank the support from the Swiss National Science Foundation and the Portuguese Funda\c{c}\~ao para Ci\^encia e Tecnologia in the form of a scolarship.
                                                                               
\end{acknowledgements}

\end{document}